# Colossal Electroresistance and Colossal Magnetoresistance in Spinel Multiferroic CdCr$_2$S$_4$

C. P. Sun[1], C. L. Huang[1], C. C. Lin[1], J. L. Her[1], C. J. Ho[1], J. -Y. Lin[2], H. Berger[3], and H. D. Yang[1*]

[1]Department of Physics, Center for Nano Science and Nano Technology, National Sun Yat-Sen, University, Kaohsiung 804, Taiwan
[2]Institute of Physics, National Chiao-Tung University, Hsinchu 300, Taiwan
[3]Institutes of Physics of Condensed Material, EPFL, Lausanne, Switzerland

Colossal magnetoresistance (CMR) and electroresistance (CER) induced by the electric field in spinel multiferroic CdCr$_2$S$_4$ are reported. It is found that a metal-insulator transition (MIT) in CdCr$_2$S$_4$ is triggered by the electrical field. In magnetic fields, the resistivity $\rho$ of CdCr$_2$S$_4$ responds similarly to that of CMR manganites. Combing previous reports, these findings make CdCr$_2$S$_4$ the unique compound to possess all four properties of the colossal magnetocapacitive (CMC), colossal electrocapacitive (CEC), CER, and CMR. The present results open a new venue for searching new materials to show CMR by tuning electric and magnetic fields.

The discovery of colossal magnetoresistance (CMR) and the efforts to understand it are among the most sensational events of the past few decades in the field of strongly correlated electron systems.[1] CMR in manganites is thought to be relevant to the double exchange (DE) mechanism,[2] Jahn-Teller (JT) distortion[3] and correlated polarons (CP)[4]. For a number of years, research efforts have accumulated an enormous amount of knowledge on CMR manganites.[3,4] In particular, these materials have provided an important venue to study charge-spin-lattice interactions. Meanwhile, *multiferroics*, the coexistence of ferromagnetic, ferroelectric and ferroelastic phases in a material, has been one of the most attractive issues in fundamental and applied researches recently.[5] Multiferroics was also claimed to exist in the chromium spinel materials CdCr$_2$S$_4$, in which ferromagnetic and ferroelectric transitions were found at 85 and 57 K, respectively.[6-8] Previous studies have shown that spinel CdCr$_2$S$_4$ have colossal magnetocapacitive (CMC)[6-8] and colossal electrocapacitive (CEC)[8] effects. Both the cubic chromium spinels CdCr$_2$S$_4$ and HgCr$_2$S$_4$ are known to show relaxor ferroelectric behavior.[6] The reported colossal magnetocapacitive (CMC) effect in these spinels has stimulated debates on the origin of the observed relaxor ferroelectric behavior.[7] Cl impurity accumulated during crystal growth and defect might be latent reason related to the observed phenomena. CMR and multiferroics were two seemingly non-correlated properties until the unexpected discovery of CMR in HgCr$_2$S$_4$.[9] In the spinels, the origin of CMR has remained elusive. HgCr$_2$S$_4$ is an antiferromagnetic spinel,[10] and therefore seems an unlikely material to show CMR. However, no phonon splitting was observed at the magnetic phase transition in HgCr$_2$S$_4$ either by Raman or far-infrared spectroscopy.[11,12] Those studies have demonstrated that HgCr$_2$S$_4$ is dominated by ferromagnetic exchange as the FM CdCr$_2$S$_4$ is. Furthermore, strong spin-phonon coupling is indicated in both HgCr$_2$S$_4$ and CdCr$_2$S$_4$. Moreover, the strong coupling between charges and spins is shown in the vicinity of the FM ($T_C$ ~85 K) and ferroelectric ordering temperatures ($T_p$~56 K) through the colossal electric field effect (EFE) on the dielectric properties in CdCr$_2$S$_4$.[8] Inspired by this recent observation, we revisit the issue of the transport property of CdCr$_2$S$_4$ using EFE. Previously, the electrical transport properties of CdCr$_2$S$_4$ was thought to be insulating as advocated by the recent LSDA+$U$ calculations.[13] Amazingly, we found that the metal-insulator transition in CdCr$_2$S$_4$ was triggered by the electric field near ferromagnetic transition. These results suggest that CMR is common for the chromium spinels that fall around the boundary between the ferromagnetism (FM) and FM fluctuations.[10] According to the previous dielectric[8] and present transport results, the applied electric field plays an important role in enhancing the spin-charge coupling in CdCr$_2$S$_4$.

Single crystal CdCr$_2$S$_4$ was grown by chemical transport reaction, using chlorine as the transport agent. Detail information can be found in Ref. [8]. Temperature and field dependent magnetization was measured by SQUID XL-7 magnetometer. High accuracy AC calorimetry with chopped light as a heat source was adapted to measure temperature dependent specific heat.[14] The temperature-dependent resistivity measurements were performed with various magnitudes of the driving currents and under magnetic fields to study the electrical transport properties of CdCr$_2$S$_4$. Conventional four-point method was used to measure the resistivity of single crystal CdCr$_2$S$_4$. A high resistance sourcemeter was integrated into PPMS (Physical Properties Measurement System) for resistivity measurements.

Fundamental physical properties of the single crystal used in the present work are shown in Fig. 1. The magnetization and specific heat indicate the presence of FM with an ordering temperature $T_C$ ~ 85 K, which are consistent with those reported by Hemberger *et. al.*.[6] To avoid the concern of sample quality under high current measurement, the properties of magnetization and specific heat were checked and confirmed to be identical as initial. The temperature dependence of resistivity with various driving currents is shown in Fig. 2. The semiconducting behavior from applying a small DC current is also consistent with LSDA+$U$ calculation[13] and the earlier electrical transport measurements.[15] However, with an increase in the applied current, an astonishing feature appears near $T_C$. A clear metal-insulator transition (MIT) near $T_C$ was observed in CdCr$_2$S$_4$, mimicking the electric transport property of CMR manganites. The peak temperature (or the metal-insulator transition temperature) $T_{MIT}$ remains at the same temperature as $T_C$ with the increasing current. It is therefore tempting to consider the magnetic origin of this MIT in CdCr$_2$S$_4$. Indeed, a mechanism of spin-charge coupling is suggested by further experiments as the fol In the inset of Fig.2, the temperature hysteresis of resistivity is shown that no charge-ordering behavior which is commonly observed in the CMR-manganite.lowing. Moreover, the conductivity is enhanced by EFE throughout the entire temperature range. Consequently, both the electric field and the magnetic field effects on the electrical transport properties can be observed in $CdCr_2S_4$.

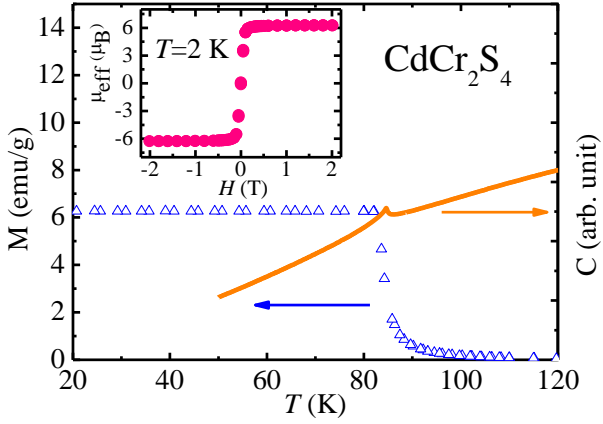

FIG.1. Temperature dependent magnetization and specific heat of $CdCr_2S_4$. Saturation moment of $6\mu_B$ is derived from the isothermal magnetization at 2 K as shown in the inset.

To see if this EFE-induced MIT is really similar to CMR in manganites, the magnetic field $H$ is applied. As a result, $CdCr_2S_4$ clearly displays CMR as shown in Fig. 3. $T_{MIT}$ can be elevated to higher temperatures by increasing $H$. To further establish the existence of the charge-spin coupling, both $T_{MIT}$ and $T_C$ were plotted in the inset of Fig. 3. The correlation between $T_{MIT}$ and $T_C$ can be vividly confirmed. To quantify the EFE-induced CER and CMR, the difference of the resistance with changes in $E$ and $H$ are shown in Fig. 4, where the ratio is defined as
$$\frac{R(E_f, H_f, T) - R(E_i, H_i, T)}{R(E_i, H_i, T)} \times 100\%.$$
For ER ratio, $I_f$ and $I_i$ are 500 nA and 2.5 nA, respectively, with $H$=0 T. For MR ratio, $H_f$ and $H_i$ are 9 T and 0 T, respectively, with $I$=10 nA. This clearly shows that the ER ratio is quite sensitive to the driving current and significantly increases below $T_C$. The MR ratio near $T_C$ is as high as 55 %. It is noted that the CMR magnitude in $CdCr_2S_4$ is slightly smaller than in manganites. However, it is comparable to that in $HgCr_2S_4$, the mechanism of CMR inside is not clearly understood at present.[9]

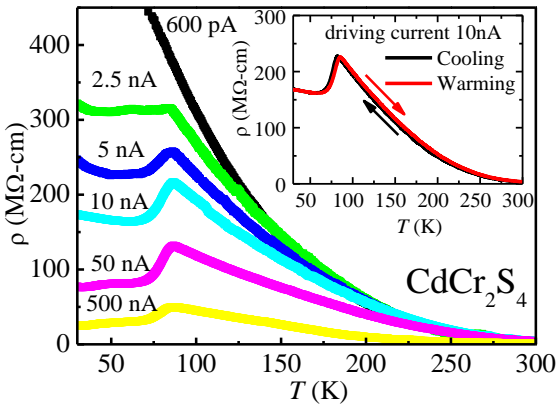

FIG.2. The insulating behavior is observed with small driving current. With increasing applied current from 600 pA to 500 nA, strong suppression of resistivity and MIT around $T_C$ are clearly observed in multiferroic spinel $CdCr_2S_4$. Inset shows no temperature hysteresis of resistivity.

Thus, both CER and CMR exist in $CdCr_2S_4$. These new findings certainly deserve further consideration, even there are few concerns on the impurity and defect level. CMR is mostly found in manganites, which involves DE due to mixed valance Mn ions and JT distortion polarons or CP. In the ferromagnetic semiconductors, another mechanism due to the magnetic polarons could lead to CMR.[16-18] It is plausible that CMR of $CdCr_2S_4$ (also a ferromagnetic semiconductor) is due to the latter mechanism. Indeed, the origin of CMR in the similar systems of Cd-doped $FeCr_2S_4$ and $HgCr_2S_4$ has been discussed in the context of magnetic polarons.[9,18] Comparing between Fe-doped $CdCr_2S_4$ and $CdCr_2S_4$ with EFE in the present study, we find that EFE is very similar to Fe doping effects, both of which induce MIT and CMR in $CdCr_2S_4$. The existence of magnetic polarons in Cd doped $FeCr_2S_4$ has been previously reported.[18] It is also known that Fe doping in $CdCr_2S_4$ leads to an increase in the carrier density $n$. When the density of magnetic polarons reaches $n=1/\xi^3$, where $\xi$ is the FM correlation length, the carriers become itinerant and MIT occurs. It is very likely that EFE on $CdCr_2S_4$ increases $n$ as Fe doping does. The possible mechanism could be that EFE leads to further local lattice distortion[11,12] and consequently increases $n$. In the context of magnetic-polaron-induced MIT, a tiny change of $n$ should suffice, as in the case of $Cd_{1-x}Fe_xCr_2S_4$.[18] Further careful examination on the inset of Fig. 3 reveals that $T_{MIT}$ is indeed slightly higher than $T_C$ by $T_{MIT} \approx 1.05 \sim 1.1\ T_C$, showing nice consistence with the predication of the magnetic polaron model.[19,20] Since $\xi$ diverges as $T$ approaches $T_C$ from the high temperatures, the condition of $n=1/\xi^3$ is reached for $T>T_C$. The experimental results are in accord with this fundamental reasoning. The values of $T_{MIT}/T_C \approx 1.05 \sim 1.1$ further imply $J'/t \approx 1$ as in Mn pyrochores,[16] where $t$ is hopping constant and $J'$ is the effective exchange coupling between the ion spins and the carriers. Since the relation of $n=1/\xi^3$ is likely a boundary of crossover, the transition is not sharp as observed in Figs. 2 and 3.

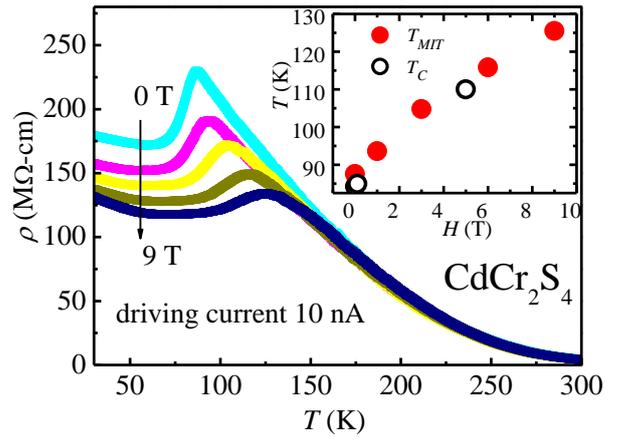

FIG.3. CMR is clearly shown at the fixed driving current (10 nA), with the external magnetic fields $H$=0, 1 T, 3 T, 6 T, and 9 T. In the inset, $T_{MIT}$ and $T_C$ were plotted to compare two characteristic temperatures from the transport and magnetic measurements, respectively.

It seems unlikely to ascribe CMR in $CdCr_2S_4$ to DE mechanism, as the valence of Cr is generally believed to be very close to 3+ from magnetization [see Fig. 1]. Nevertheless, an earlier report supported the mixed valence of Cr in $CdCr_2S_4$ derived from ferromagnetic resonance experiments.[21] This spectroscopy result is not inconsistent with the magnetization (~6$\mu_B$), where the saturation moment is for $Cr^{3+}$, if the orbital angular moment of the Cr ion is not completely quenched. Even though, the present spectroscopic data do not advocate DE mechanism which requires the mixed valence of Cr. On the other hand, this non-ionic Cr-S bond nature is further in accord with the results from far-infrared spectroscopy.[11] It is intriguing that the Cr-S bond in $CdCr_2S_4$ is more ionic than in $HgCr_2S_4$.[11] It cannot be ruled out that EFE could lower the ionicity of Cr-S bond in $CdCr_2S_4$ to the level of that in $HgCr_2S_4$ and induce MIT. As mentioned above, lattice distortion as another important ingredient for CMR in manganites has been reported in the literature. However, $T_C$ in $CdCr_2S_4$ decreased with applying pressure ($dT_C/dP<0$).[22,23] This pressure effect is dissimilar to that observed in manganese oxides with DE interaction.[24] The chromium spinels have a very low $n$, considered from electrical transport measurement. Whether the carrier hopping via DE could lower enough kinetic energy and lead to MIT certainly remains another open question.

In addition to CMR effect, EFE behavior in $CdCr_2S_4$ is very similar to that in $HgCr_2S_4$ in many ways. For example, EFE enhances the dielectric constant of $CdCr_2S_4$ at low temperatures by one order of magnitude,[8] reaching the same value of $HgCr_2S_4$.[9] CMC with EFE in $CdCr_2S_4$ is also closer to that in $HgCr_2S_4$ than in $CdCr_2S_4$ without applied EF. This similarity strongly implies that CMR or CMC in both $CdCr_2S_4$ with EFE and $HgCr_2S_4$ share the same origin. Local lattice distortion through exchangestriction and EFE is certainly a viable candidate. Therefore, temperature dependent microscopic technique might be helpful to elucidate the present issue. It is noted that the EF-induced MIT by the interplay of the ionic and electric conduction has been reported recently in multiferroics among promising applications.[25]

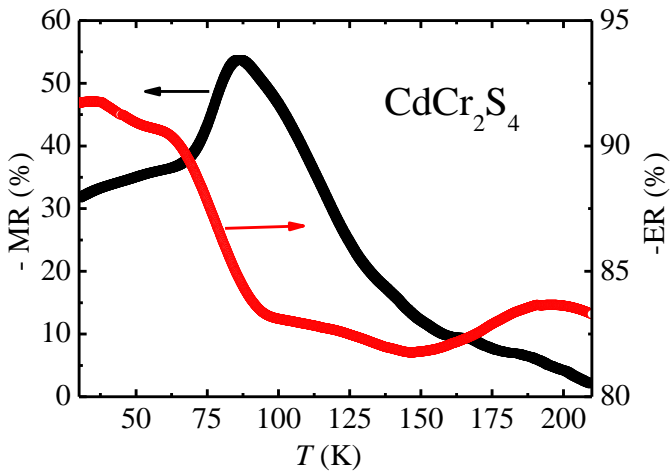

FIG.4. The ER ratio is increased rapidly below $T_C$, where driving current $I_f$ and $I_i$ are 500 nA and 2.5 nA, respectively with $H=0$. The MR ratio near $T_C$ is the largest and up to 55 %, where $H_f$ and $H_i$ are 9 T and 0 T, respectively with $I=10$ nA.

In summary, we observe EF-induced MIT, the discovery of CMR induced by the external field is reported in the geometrically frustrated spinel $CdCr_2S_4$. EFE induced CMR effect is unique and provides a novel venue for the interplay of electric and magnetic fields. This finding makes $CdCr_2S_4$ the only material known to show CMC, CEC, CER and CMR altogether. The mechanism of CMR in $CdCr_2S_4$ is compared to that in $La_{1-x}Ca_xMnO_3$ and turns out to be different. The present results open a new venue for searching new materials to show CMR by tuning electric and magnetic fields.

This work was supported by National Science Council of Taiwan under grant No. NSC 97-2112-M110-005-MY3. H. B. thanks the NCCR research pool MaNEP of the Swiss National Science Foundation for support in sample preparation.

*Corresponding author: yang@mail.phys.nsysu.edu.tw
1. A. P., Ramirez, J. Phys. Condensed. Matter **9,** 8171 (1997).
2. C. Zener, Phys. Rev. **82,** 403 (1951).
3. A. J. Millis, B. I. Shraiman, and R. Mueller, Phys. Rev. Lett. **77,** 175 (1996).
4. C. Sen, G. Alvarez, and E. Dagotto, Phys. Rev. Lett. **98,** 127202 (2007).
5. W. Eerenstein, N. D. Mathur, and J. F. Scott, Nature (London) **442,** 759 (2006).
6. J. Hemberger, P. Lunkenheimer, R. Fichtl, H.-A. Krug von Nidda, V. Tsurkan, and A. Loidl, Nature **434,** 364 (2005).
7. G. Catalan and J. F. Scott, Nature **448,** E4 (2007).
8. C. P. Sun, C. C. Lin, J. L. Her, C. J. Ho, S. Taran, H. Berger, B. K. Chaudhuri, and H. D. Yang, Phys. Rev. B **79,** 214116 (2009).
9. S. Weber, P. Lunkenheimer, R. Fichtl, J. Hemberger, V. Tsurkan, and A. Loidl, Phys. Rev. Lett. **96,** 157202 (2006).
10. T. Rudolf, C. Kant, F. Mayr, J. Hemberger, V. Tsurkan, and A. Loidl, New J. phys. **9,** 76 (2007).
11. T. Rudolf, C. Kant, F. Mayr, J. Hemberger, V. Tsurkan, and A. Loidl, Phys. Rev. B **76,** 174307 (2007).
12. V. Gnezdilov, P. Lemmens, Y. G. Pashkevich, P. Scheib, C. Payen, K. Y. Choi, J. Hemberger, A. Loidl, and V. Tsurkan, cond-mat/0702362.
13. C. J. Fennie, and K. M. Rabe, Phys. Rev. B **72,** 214123 (2005).
14. Y. K. Kuo, C. S. Lue, F. H. Hsu, H. H. Li, and H. D. Yang, Phys. Rev. B **64,** 125124 (2001).
15. H. W. Lehmann, and M. Robbins, J. Appl. Phys. **37,** 1389 (1966).
16. B. Martinez, R. Senis, J. Fontcuberta, X. Obradors, W. Cheikh-Rouhou, P. Strobel, C. Bougerol-Chaillout , and M. Pernet, Phys. Rev. Lett. **83,** 2022 (1999).
17. Z. Yang, S. Tan, Z. Chen, and Y. Zhang, Phys. Rev. B **62,** 13872 (2000).
18. Z. R. Yang, X. Y. Bao, S. Tan, and Y. H. Zhang, Phys. Rev. B **69,** 144407 (2004).
19. P. Majumdar, and P. Littlewood, Phys. Rev. Lett. **81,** 1314 (1998).
20. P. Majumdar, and P. B. Littlewood, Nature **395,** 479 (1998).
21. K. G. Nikiforrov, A. G. Gurevich, S. I. Radautsan, L. M. Emiryan, and V. E. Tezlevan, Phys. Stat. Sol. **72,** K37 (1982).
22. Vishun C. Srivastava, J. Appl. Phys. **40,** 1017 (1969).
23. C. P. Sun, and H. D. Yang, unpublished.
24. C. F. Chang, P. S. Chou, L. H. Tsay, S. S. Weng, S. Chatterjee, H. D. Yang, R. S. Liu, and W. H. Li, Phys. Rev. B **58,** 12224 (1998).
25. C.-H Yang, J. Seidel, S. Y. Kim, P. B. Rossen, P. Yu, M. Gajek, Y. H. Chu, L. W. Martin, M. B. Holcomb, Q. He, P. Maksymovych, N. Balke, S. V. Kalinin, A. P. Baddorf, S. R. Basu, M. L. Scullin, and R. Ramesh, Nature Materials **8,** 485 (2009)